\documentclass[10pt]{IEEEtran}

\usepackage{enumerate} 
\usepackage{geometry}
\usepackage{graphicx,amsmath,amsfonts}
\usepackage[]{algorithm}
\usepackage{url}
\usepackage{algpseudocode}

\usepackage{tabularx}
\usepackage{array, booktabs}
\usepackage{graphicx,amsmath,amsfonts}
\usepackage{wrapfig}
\usepackage{geometry}
\usepackage{subfig} 
\usepackage{caption}
\usepackage{epstopdf}
\usepackage{cite}
\usepackage{setspace}
\usepackage{graphics}
\begin{document}

\title{Frequent Pattern Mining approach to Image Compression}

%
%

\author{\IEEEauthorblockN{Avinash Kadimisetty\IEEEauthorrefmark{1},
C. Oswald\IEEEauthorrefmark{2} and B. Sivaselvan\IEEEauthorrefmark{3} \\}
\IEEEauthorblockA{Department of Computer Engineering,
IIITDM Kancheepuram, Tamil Nadu, India\\
Email: \IEEEauthorrefmark{1}coe12b009@iiitdm.ac.in,
\IEEEauthorrefmark{2}coe13d003@iiitdm.ac.in,
\IEEEauthorrefmark{3}sivaselvanb@iiitdm.ac.in}}

\maketitle
\section{Abstract}
The paper focuses on Image Compression, explaining efficient approaches based on Frequent Pattern Mining(FPM). The proposed compression mechanism is based on clustering similar pixels in the image and thus using cluster identifiers in image compression. Redundant data in the image is effectively handled by replacing the DCT phase of conventional JPEG through a mixture of $k$-means Clustering and Closed Frequent Sequence Mining. To optimize the cardinality of pattern(s) in encoding, efficient pruning techniques have been used through the refinement of Conventional Generalized Sequential Pattern Mining(GSP) algorithm. We have proposed a mechanism for finding the frequency of a sequence which will yield significant reduction in the code table size. The algorithm is tested by compressing benchmark datasets yielding an improvement of 45\% in compression ratios, often outperforming the existing alternatives. PSNR and SSIM, which are the image quality metrics, have been tested which show a negligible loss in visual quality.  \\

\textbf{Keywords:} Closed Frequent Sequence, Clustering, Image Quality, Lossy Compression, Compression Ratio.

\section{Introduction}
 As the amount of data exchanged on the Internet is increasing enormously, the need for compressing these large volumes of data is apparent. Data exists in various forms like Text, Images, Audios, Videos etc. The transfer speed of the data is affected by its size and hence reduction in the size is preferred. Image compression is defined as the process of reducing the data required to represent the image and is based on the fact that neighbouring pixels are highly correlated, referred to as the spatial redundancy~\cite{Salomon:2004:DCC}. Image compression techniques are broadly classified as lossless and lossy. Lossless Image compression reduces the size of the image by removing or reducing the redundancy in the data. Compression techniques like Shannon-Fano Encoding, Huffman Encoding, Arithmetic Encoding, Lempel-Ziv-Welch(LZW), Bit-Plane Coding and Lossless Predictive Coding perform lossless compression~\cite{Salomon:2004:DCC, shannon1949communication, huffman1952method, witten1987arithmetic}. Lossy compression on the other hand compresses the image by removing or reducing the irrelevancy in addition to redundancy, thereby achieving better compression than lossless compression. Compression techniques like JPEG(Joint Photographic Experts Group), JPEG 2000, Wavelength Coding, Transform Coding and Lossy Predictive Coding, GIF(Graphics Interchange Format), etc perform the task of Lossy Image compression~\cite{Salomon:2004:DCC, Wallace1991,  jpeg2000skodras}. 

Data Mining is the process of extracting hidden and useful information from large DB's~\cite{HanK2000}. In Data Mining, five perspectives were observed by Naren Ramakrishnan et. al. which are Compression, Search, Induction, Approximation and Querying~\cite{RamakrishnanG99}. The perspective of Data mining as a compression technique is of interest in this study. The process of data mining focuses on generating a reduced(smaller) set of patterns(knowledge) from the original DB, which can be viewed as a compression mechanism. A few of the mining techniques such as Classification, Clustering, Association Rule Mining(ARM) may be explored from the compression perspective~\cite{HanK2000}. In this study, the knowledge of Clustering and Closed Frequent Sequence Mining(CFSM) is used to achieve efficient compression.

A formal definition of Data Clustering is as follows. Let us assume a data set, $D$, containing $n$ objects in the Euclidean space. $k$ clusters, $C_{1},\dots,C_{k}$, where $C_{i} \subset D$, $C_{j} \subset D$ and $C_{i} \cap C_{j} = \emptyset$ for $1 \leq i, j \leq k$ are distributed by the objects in $D$ with the help of the partitioning methods~\cite{HanK2000, Jain2010kmeans}. To assess the quality of partitioning, an objective function is used such that high similarity is observed in the objects within the same cluster and high dissimilarity is observed with respect to the objects in other clusters. Thus the objective function aims to maximize intra cluster similarity and minimize inter cluster similarity. Clustering algorithms can be classified based on Centroid, Hierarchical, Distribution and Density.  The well known Data Clustering algorithms include \textit{k-means}, PAM, CLARA, DIANA, AGNES, DBSCAN, BIRCH, OPTICS etc \cite{HanK2000, Jain2010kmeans, pamkaufman, clarans2002,  ester1996density, birch1996, ankerst1999optics}.

Let $S$ be a set of items. A set $X$ = $\{a_1, a_2, \dots, a_l\} \subseteq S$ is called an itemset or \textit{l}-itemset if it contains \textit{l} items. A sequence \textit{s} is a tuple $s = \langle S_1 S_2 S_3 \dots S_x\rangle$ with $S_i \in S$, and $\forall i: 1 \leq i \leq x$. The size of a sequence is denoted as $|s|$, the number of itemsets in that sequence. Every sequence with \textit{l} items is a called an \textit{l}-sequence. We say the sequence $\beta = \langle S_{a_1} S_{a_2} \dots S_{a_x}\rangle$ is a subsequence of another sequence $\gamma = \langle S_{b_1} S_{b_2} \dots  S_{b_y}\rangle$ (or $\gamma$ is a supersequence of $\beta$), denoted $\beta \subseteq \gamma$, if there exists integers $1 \leq i_1 < i_2 < \dots < i_ x \leq y$ such that $S_{a_1} \subseteq S_{b_{i_1}}$, $S_{a_2} \subseteq S_{b_{i_2}}$, \dots, $S_{a_x} \subseteq S_{b_{i_x}}$\cite{Gomariz2013}. In the context of this work, a sequence is considered to be the characters in an ordered and contiguous manner w.r.t the index. Consider the text `\textit{abracadabra}'\ where `\textit{ara}'\ is a sequence in conventional terms but in our work, `\textit{abra}'\ is only considered as a sequence. The terms sequence and pattern are used interchangeably in the rest of this article. 

\vspace*{0.8cm}

\subsection{Related Work}

Lossy image compression techniques include Chroma Subsampling, Transform Coding, Fractal Compression, Vector Quantization, Block Truncation etc~\cite{Winkler2001, fisher1994fractal, gersho2012vector, delp1979image}. Chroma Subsampling is based on the fact that human eye has lower acuity for color differences. Chroma Subsampling holds intensity information and color information of lower resolution. Transform Coding produces an output of low quality of the original image. In Transform Encoding, the input image is split into 8 $\times$ 8 blocks and then DCT is applied to each and every block. Quantization table is used to compress each block. The compressed image can be formed with all the compressed blocks. During Decompression, Inverse DCT is used. Fractal Compression is used to encode images into mathematical data and describe the fractal properties of the image. In Fractal compression, the input image is split into Parent Blocks and each parent block is split into Child Blocks. Each child block is compared against a subset of all possible overlapping blocks of parent block size. The parent block for which the lowest difference occurs is identified  according to a pre defined measure. Grayscale transform is calculated to match the intensity levels of parent block and child block. In Vector Quantization, the encoder looks for a closest match codeword for every image block in the dictionary and the index is sent to the decoder. The decoder replaces the index with the corresponding codeword from the dictionary. Block Truncation is a simple and fast compression technique. The key concept is to perform moment preserving quantization for blocks of pixels. Recent works in Image Compression include Clustering based,  Segmentation based and Fractal compression based approaches. Clustering based Image compression algorithms filter the image to reduce the noise and perform Fuzzy image enhancement. The image is split into square blocks and using DCT each square block is transformed. Using DCT cosine membership functions, membership values and cluster centroids are found to create a segmented image and perform one dimensional run length encoding. Some of the demerits of the above discussed clustering based techniques include noticeable reduction in quality, overhead in the compressed size leading to a low compression ratio.

\vspace*{-0.3cm}

\section{Proposed Image Compression Algorithm}
\subsection{Encoder}


\begin{algorithm}[h]
\caption{\textbf{Image Compression}}\label{euclid}
\begin{algorithmic}
\State \textbf{Input:}{
	\begin{enumerate}
		\item Image \textit{$I_{[m\times n\times3]}$} in RGB color space.
		\item Number of clusters \textit{k} in \textit{k-means} algorithm
 		\item Minimum support $\alpha$ in GSP algorithm
	\end{enumerate}
}
\State \textbf{Output: }{The compressed image $I_c$}
\State Split \textit{I} into Red($I_1$), Green($I_2$) and Blue($I_3$) components\;
\State  \textit{i} $\gets$ 1\;
\While{$i \leq$ 3}
\State 	$C_i$ = \textit{k-means\_Clustering($I_i, k$)}; \Comment $C_i$ is the cluster identifier table for each component $I_i$
\State 	Replace each pixel in $I_i$ with the corresponding cluster identifier from $C_i$ to get $I_i^{'}$;
\State	$S_{i}$ = \textit{GSP\_Modified}($I_i^{'},\:\alpha$); 
\State 	$S_{i}^{'}$ = \textit{Modified\_Support}($S_{i}, I_{i}^{'}$); 
\State	$T_i$ = \textit{Huffman\_Encoding}($S_{i}^{'}$);
\State 	For each sequence $\sigma$ in $S_{i}^{'}$, replace $\sigma$ in $I_i^{'}$ with the corresponding Huffman code from $T_i$ to get the compressed image $I_c$;
\State 	\textit{i++};
\EndWhile
\end{algorithmic}
\end{algorithm}

\subsubsection{Image Acquisition}

Overall approach of our proposed work is mentioned in Algorithm 1. The source image \textit{I} is chosen in RGB color space and is converted to an \textit{m}$\times$\textit{n}$\times$\textit{3} matrix of resolution \textit{m}$\times$\textit{n}. The image \textit{I} is split into three components $I_1,\: I_2,\: I_3$ corresponding to the Red, Green and Blue components. Each element in the matrices $I_1,\: I_2,\: I_3$ represents a pixel value. Every pixel can be represented with \textit{k}-bits if the pixel value lies in [0, $2^k-1$]. The image matrix \textit{B} is considered for explaining the working of the algorithm. 

\[
B = 
\begin{bmatrix}
    154 & 123 & 123 & 123 & 123 & 123 & 123 & 136 \\
    192 & 180 & 136 & 154 & 154 & 154 & 136 & 110 \\
    254 & 198 & 154 & 154 & 180 & 154 & 123 & 123 \\
    239 & 180 & 136 & 180 & 180 & 166 & 123 & 123 \\
    180 & 154 & 136 & 167 & 166 & 149 & 136 & 136 \\
    128 & 136 & 123 & 136 & 154 & 180 & 198 & 166 \\
    123 & 105 & 110 & 149 & 136 & 136 & 180 & 166 \\
    110 & 136 & 123 & 123 & 123 & 136 & 154 & 136 
\end{bmatrix}
\]

\subsubsection{Image Clustering}

In \textit{k-means} clustering, \textit{n} objects are partitioned into \textit{k} clusters, such that each object belongs to the cluster with the nearest mean~\cite{Jain2010kmeans}. In the context of Image, each pixel value is treated as the data object. When Clustering is performed on the image with \textit{k} clusters using \textit{k-means} algorithm, the pixels are grouped based on their relative closeness. Every cluster is represented by a cluster identifier and each cluster identifier is associated with a pixel value which is the cluster center(mean of all the pixels in that cluster). If $|C|$ and $|P|$ denote the number of bits required to represent the cluster identifier and the pixel respectively, then $|C| \leq |P|$. Thus every pixel in each component of the image \textit{I} is replaced with the cluster identifier of the cluster to which the pixel belongs, to get the component matrices $I_{1}^{'}, \: I_{2}^{'},\: I_{3}^{'}$. Clustering is performed on $B$ with $k=5$ and the cluster identifiers are shown in table I. $B^{'}$ is formed from $B$ by replacing each pixel value with the corresponding cluster identifier.

\begin{table}[!htb]
\begin{minipage}{0.5\textwidth}
\centering
	\begin{tabular}{c c c}
		\hline
		Cluster Identifier & Mean Pixel value \\
		\hline
		0 & 179 \\
		1 & 153 \\
		2 & 135 \\
		3 & 246 \\
		4 & 120 \\
		\hline
	\end{tabular}
	\caption{Cluster Identifier table of \textit{B} with \textit{k} = 5}
\end{minipage}
\begin{minipage}{0.5\textwidth}
\[B\ensuremath{'} = \begin{bmatrix}
    1 & 4 & 4 & 4 & 4 & 4 & 4 & 2 \\
    0 & 0 & 2 & 1 & 1 & 1 & 2 & 4 \\
    3 & 0 & 1 & 1 & 0 & 1 & 4 & 4 \\
    3 & 0 & 2 & 0 & 0 & 0 & 4 & 4 \\
    0 & 1 & 2 & 0 & 0 & 1 & 2 & 2 \\
    2 & 2 & 4 & 2 & 1 & 0 & 0 & 1 \\
    4 & 4 & 4 & 1 & 2 & 2 & 0 & 0 \\
    4 & 2 & 4 & 4 & 4 & 2 & 1 & 2
\end{bmatrix}\]
\end{minipage}
\end{table}

\subsubsection{Closed Frequent Sequence Mining}

The support $\alpha$,  of a sequence $\sigma$, is defined as $|X|$ where $X = \{Y|\sigma \subseteq Y \: \land \: Y \in D\}$. 
If a sequence satisfies a user specified threshold \textit{min\_sup} denoted as minimum support $\alpha$, then it is a frequent \textit{pattern/sequence}. A frequent sequence $\sigma$ is considered to be a \textit{closed} frequent sequence if there is no other supersequence of $\sigma$ with the same support~\cite{yan2003clospan}. Otherwise, if $\gamma \supseteq \sigma$ and if \textit{support}($\sigma$) = \textit{support}($\gamma$), then we call $\sigma$, a non-closed sequence and $\gamma \:\:  absorbs  \:\: \sigma$. If $|CF|$ denotes the cardinality of closed frequent sequences and $|F|$ denotes the cardinality of frequent sequences, then $|CF| \leq |F|$.

\begin{algorithm}[!htb]
\caption{\textbf{Closed Frequent Sequence Mining}}\label{euclid}
\begin{algorithmic}
\State \textbf{Input:} {A Sequence database $D$ and $min\_sup\:\alpha$}
\State \textbf{Output: }{The Set of all Closed Frequent Sequences \textit{S}}
\Procedure{\textit{GSP\_Modified}}{}
\State $E_1$ $\gets$ $\{x|x \in D \wedge |x|=1\}$
\State $E_2$ $\gets$$E_1$$\times$$E_1$;
\State $\forall \gamma$ in $D$, for $\sigma$ in $E_2$, \textit{support}($\sigma)$++ if $\gamma \supseteq \sigma$;
\State $F_2 \gets$ \{$a \in E_2$ $\ni$ $support(a) \geq \alpha $\};
\State $k \gets 3$;
\While{$F_{k-1}$!=NULL}
\State $E_k$ = $F_{k-1} \times F_{k-1}$  \Comment If $a,b \in F_{k-1}$ and if \textit{(k-2)} length suffix of $a$ == \textit{(k-2)} length prefix of $b$ then $a+b[k-1] \in E_k$
\For{each $\sigma$ in $E_k$}
\If{support($\sigma$) $\geq \alpha$}
\State $F_k \gets F_k \cup \sigma$;
\EndIf
\EndFor
\For{each $\sigma$ in $F_k$}
\For{each $\gamma$ in $F_{k-1}$}
\If{$\gamma \subseteq \sigma$ and \textit{support}($\sigma$)==\textit{support}($\gamma$)}
\State $F_{k-1} \gets F_{k-1} - \gamma$;
\EndIf
\EndFor
\EndFor
\State \textit{k++;}
\EndWhile
\State $S = E_1 \cup F_2 \cup F_3 \cup \dots \cup F_{k-1}$;
\EndProcedure
\end{algorithmic}
\end{algorithm}

Algorithm 2 explains about the working of our modified closed frequent sequence mining approach to Image compression with a trace from table II. Each component of $I^{'}$ is treated as a sequence database $D$ in which each row is a sequence. Closed sequence patterns can be mined from a sequence database in two ways. 1) All frequent sequences are mined and a post processing step is executed  to prune out all the non closed frequent sequences. 2) Mine the closed sequences level wise by eliminating the non closed sequences \cite{Gomariz2013}. Method 2 is followed in this paper. We use Generalized Sequential Pattern(GSP) Mining \cite{srikant1996mining} algorithm for this purpose. GSP algorithm is a level wise approach to mine frequent sequences. At each level, we execute an additional step to prune out all non-closed sequences which assures that only closed frequent sequences are carried forward. Candidate set $E_i$, at level \textit{i}, is considered to be the set of all sequences at level $i$. For every sequence $s \in E_i$, if $support(s)\geq \alpha,\: s$ is a frequent sequence. The set of all frequent sequences constitute the frequent sequence set.


There may be some character/s which may not occur as subpatterns(subsequences) in any of the patterns generated for encoding. This may lead to an ambiguous encoding leaving those characters without getting encoded. In order to overcome this issue, we keep all patterns in $E_{1}$. There are several sequences whose sub sequences are also frequent. For example, the sequences \textit{444} and \textit{44} with support $3$ and $5$ are frequent. As the sequence \textit{444} includes $3$ occurrences of \textit{44}, the count of \textit{44} needs to be updated. If this overlapping count is not updated, this will increase the cardinality of $S$ in which many of the patterns which have this overlap count will not used in the encoding process. This in turn will increase the code table size tremendously, leading to a low compression ratio. Let us consider $H$ to be set of all lengthier sequences and $L$ be the set of proper subsequences from $S$. If the overlapping frequency of a sequence in $L$ is not updated, sequences in $L$ which actually have lesser support than a sequence in $H$ will pretend to have high frequency. As the count of these sub-sequences is already counted in its supersequences in $H$, as already explained, the support of these sub-sequences needs to be updated. Hence the need for finding the modified support($s_{mod}$) of every patterns in $S$ is apparent. Lengthier sequences in the set $H$ are given more priority for encoding than shorter length sequences in the set $L$, because of which the overlapping frequency of sequences in $L$ is reduced.

For example, let us consider $M$ as the matrix where each row is a sequence of cluster identifiers and $s_{mod}$ is calculated for sequences in $M$. The modified support(${s}_{mod}$) of a sequence ${s}\ensuremath{'}$ is defined as, $|\{{s}\ensuremath{'}|{s}\ensuremath{'} \subseteq M\}|$, which denotes the support of ${s}\ensuremath{'}$ in a non-overlapping manner in $M$. The patterns with modified support forms set $S\ensuremath{'}$ and $|S\ensuremath{'}| \leq |S|$. ${s}_{mod}$ eliminates the issue of overlapping pixels between sequences in $M$. In order to obtain $s_{mod}$, all the closed frequent sequences are sorted in descending order of their length. For each sequence($\sigma$) in \textit{S}, its support is obtained, after which all instances of $\sigma$ are removed from $I^{'}$ and the frequency of other sequences in $S$ is also updated. If two sequences have same frequency then priority is given to sequences with higher ASCII value. This process is done until $I^{'}$ becomes empty. The set of all closed frequent sequences along with their support counts and modified support is shown in the Table III. 

The frequent sequences and their support in $M$ are \{\textit{444} - 2, \textit{44} - 2, \textit{22} - 2, \textit{00} - 2, \textit{0} - 2, \textit{1} - 1, \textit{2} - 2 and \textit{4} - 2\}.
\[
M = 
\begin{bmatrix}
4 & 4 & 4 & 1 & 2 & 2 & 0 & 0 \\
4 & 0 & 0 & 4 & 4 & 4 & 2 & 2
\end{bmatrix}
\]

As \textit{444} is the sequence with highest length, all instances of \textit{444} are removed from $M$. 
\[
M = 
\begin{bmatrix}
  &  &  & 1 & 2 & 2 & 0 & 0 \\
4 & 0 & 0 &  &  &  & 2 & 2
\end{bmatrix}
\]
The support of all other sequences is updated. They are as follows \{\textit{444} - 2, \textit{44} - 0, \textit{22} - 2, \textit{00} - 2, \textit{0} - 2, \textit{1} - 1, \textit{2} - 2 and \textit{4} - 1\}. The process is repeated until $M$ is empty and the modified support of the sequences is found. The modified supports are \{\textit{444} - 2, \textit{44} - 0, \textit{22} - 2, \textit{00} - 2, \textit{0} - 0, \textit{1} - 1, \textit{2} - 0, \textit{4} - 1\}

\begin{algorithm}[!htb]
\caption{\textbf{Find Modified Support($s_{mod}$)}}\label{euclid}
\begin{algorithmic}
\State \textbf{Input: }{Sequences in $S$ along with their support and a Sequence Database $D$}
\State \textbf{Output: }{$S^{'}$ along with their $s_{mod}$}
\Procedure{\textit{Modified\_Frequency}}{}
\State Sort sequences in $S$ based on their length in descending order;
\For{each sequence $\sigma$ in $S$}
\State Find the frequency of $\sigma$ in $D$;
\State Remove all instances of $\sigma$ from $D$ and update frequencies of all other sequences in $S$ to get $S^{'}$;
\EndFor
\EndProcedure
\end{algorithmic}
\end{algorithm}

\begin{center}
\begin{table*}[!htb]
\centering
\begin{tabular}{c|c|c}
\hline
\textbf{Sequence} & \textbf{Candidate} & \textbf{Frequent} \\
\textbf{Length} & \textbf{Sequence Set} & \textbf{Sequence Set} \\
\hline
1 & $C_1$ = \{0(6), 1(7), 2(7), 3(2), 4(7)\} & No Pruning done \\
\hline
 & $C_2$ = \{00(5), 01(3), 02(2), 03(0), 04(1), 10(2), 11(2),   & \\ 2 &  12(4), 13(0), 14(2), 20(3), 21(3), 22(3), 23(0), 24(3), & $F_2$ = \{00(5), 01(3), 12(4), 20(3), 21(3)  \\ &  30(2), 31(0), 32(0), 33(0), 34(0), 40(0), 41(1), 42(3),  & 22(3), 24(3), 42(3), 44(5)\} \\ &  43(0), 44(5)\} &  \\
\hline
 & $C_3$ = \{000(1), 001(2), 012(1), 120(1), 121(0), 122(2), & $F_3$ = \{200(3), 444(3)\} \\
3 & 124(1), 200(3), 201(0), 212(1), 220(1), 221(0), 222(0),  &  $F_2$ = \{00(5), 01(3), 12(4), 21(3)  \\
  & 224(1), 242(1), 244(1), 420(0), 421(2), 422(0), 424(1), & 22(3), 24(3), 42(3), 44(5)\} \\
  &   442(2), 444(3)\} &  \\
\hline
4 & $C_4$ = \{4444(1)\} & $F_4$ = $\emptyset$ \\

\hline
\end{tabular}
\caption{Closed Frequent Sequence Mining on $B^{'}$ with $\alpha$ = 3}
\end{table*}
\end{center}

\begin{center}
\begin{table*}[!tbh]
\centering
\begin{tabular}{|c|c|c|c|c|c|c|c|c|c|c|c|c|c|c|c|}
\hline
Closed Frequent Sequence & 444 & 200 & 00 & 01 & 12 & 21 & 22 & 24 & 42 & 44 & 0 & 1 & 2 & 3 & 4 \\
\hline
Support & 3 & 3 & 5 & 3 & 4 & 3 & 3 & 3 & 3 & 5 & 6 & 7 & 7 & 2 & 7\\ \hline
Modified Support($s_{mod}$) & 4 & 3 & 2 & 2 & 4 & 2 & 0 & 1 & 1 & 2 & 3 & 5 & 4 & 2 & 1\\
\hline
\end{tabular}
\caption{Closed Frequent Sequences with their corresponding support and $s_{mod}$}
\end{table*}
\end{center}

\vspace*{-0.3cm}

\subsubsection{Huffman Encoding}

Huffman encoding is now performed on $S^{'}$ to generate the Huffman tree given in figure 1. The matrix $I^{'}$ is encoded using code table $T$ to generate the compressed image $I_c$ with the preference to lengthier sequences. As a pixel can be represented using 8 bits, the original image needs 8 $\times$ 8 $\times$ 8 = 512 bits to encode. The final compressed image obtained through our algorithm requires only 129 bits thereby achieving a compression ratio of 3.968.

\begin{center}
\begin{figure}
\includegraphics[width = 0.5\textwidth]{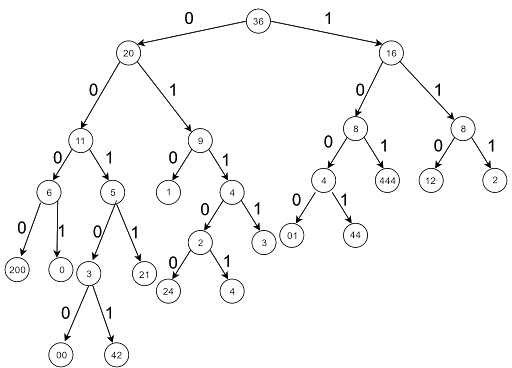}
\caption{Huffman Tree for the Closed Frequent Sequences}
\label{fig:}
\end{figure}
\end{center}

%
%

\subsection{Decoder}


\begin{algorithm}[h]
\caption{\textbf{Image Decompression}}\label{euclid}
\begin{algorithmic}
\State \textbf{Input: }{
		 Compressed Image $I_c$, Code Table $T$ and Cluster Identifier Table $C$
}
\State \textbf{Output: }{Decompressed Image $I_d$}
\Procedure{\textit{Decompress}}{}
\State Decode $I_c$ using $T$ to get $I^{'}$
\State Replace each cluster identifier in $I^{'}$ with the corresponding mean pixel value to get $I_{d}$
\EndProcedure
\end{algorithmic}
\end{algorithm}

The decoder is given in Algorithm 4 and the the decompressed image matrix for the input $B$ is given below. 
\[
I_d = 
\begin{bmatrix}
153 & 120 & 120 & 120 & 120 & 120 & 120 & 135 \\
179 & 179 & 135 & 153 & 153 & 153 & 135 & 120 \\
246 & 179 & 153 & 153 & 179 & 153 & 120 & 120 \\
246 & 179 & 135 & 179 & 179 & 179 & 120 & 120 \\
179 & 153 & 135 & 179 & 179 & 153 & 135 & 135 \\
135 & 135 & 120 & 135 & 153 & 179 & 179 & 153 \\
120 & 120 & 120 & 153 & 135 & 135 & 179 & 179 \\
120 & 135 & 120 & 120 & 120 & 135 & 153 & 135  
\end{bmatrix}
\]

\section{Simulation Results and Discussion}

The simulation is performed on an Intel core i5-4210 CPU 1.70 GHz, 4GB Main Memory and 1TB Secondary Memory and the algorithm is implemented in Python 3. The CFSM-IC algorithm is tested on various benchmark images like Lena, Peppers, Baboon  of 512$\times$512 size in \textit{bmp} format\cite{imagedataset}. The results are shown for Lena and Peppers images whose pixel values lie in the range 0 to 255. Table IV shows the various parameters for the standard images Lena, Peppers, Baboon and Boat, each of size 512$\times$512. The parameters Number of Clusters \textit{k} and Minimum Support $\alpha$ control the compression ratio and only $k$ affects the quality of the image. Hence the results are observed by varying the parameters \textit{k} and $\alpha$ to study their effects on compression. The Compression ratio $C_r$ is defined as
\begin{center}
$C_{r}$ $ = \frac{\text{\normalsize Uncompressed size of Image}}{\text{\normalsize Compressed size of Image}}$
\end{center}

\begin{center}
\begin{table*}[!tbh]
\centering
\setlength{\tabcolsep}{4pt}
\begin{tabular}{c c c c c c c c c c c c c c c}
\hline

\textbf{Image} & \textit{\textbf{k}} &  \textbf{Code } &  \textbf{Total Comp-} &  \textbf{JPEG} & \textbf{$C_r$} & \textbf{JPEG $C_r$} & \textbf{GIF $C_r$} & \textbf{PSNR} & \textbf{PSNR} &  \textbf{PSNR} \\

& & \textbf{Table } & \textbf{ressed} & \textbf{Size*} & No & No & No &\textbf{Proposed} & \textbf{JPEG} & \textbf{GIF} \\

& & \textbf{Size*} & \textbf{ Size*} & & Unit & Unit & Unit & No Unit & No Unit  & No Unit  \\ 

\hline
Lena & 8 & 15.165 &  146.781 &  404 & 5.35 & 1.94 & 3.48 & 35.85 & 53.6 & 29.55 \\
 & 15 &  9.1855 &  253.86 &  404 & 3.098 & 1.94 & 3.48 & 40.59 & 53.6 & 29.55\\
 & 24 & 5.657 &  336.157 &  404 & 2.33 & 1.94 & 3.48 & 44.02 & 53.6 & 29.55\\
Peppers & 8 & 22.835 &  172.43 &  439 & 4.56 & 1.83 & 3.76 & 34.71 & 54.06 & 29.57 \\
 & 15 & 9.459 &  277.416  & 439 & 2.83 & 1.83 & 3.76 & 39.42 & 54.06 & 29.57 \\
 & 24 & 8.09 &  365.06 &  439 & 2.15 & 1.83 & 3.76 & 43.12 & 54.06 & 29.57 \\ 
Baboon & 8 &  8.80 &  292.33 &  548 & 2.69 & 1.43 & 3.08 & 33.96 & 53.79 & 28.47 \\
 & 15 &  5.74 &  383.30 & 548 & 2.05 & 1.43 & 3.08 & 38.81 & 53.79 & 28.47   \\
 & 24 &  4.83 &  444.23 & 548 & 1.77 & 1.43 & 3.08 & 42.08 & 53.79 & 28.47  \\
Boat & 8 &  10.44 &  205.28 &  481 & 3.83 & 1.63 & 3.34 &  34.77 & 53.85 & 28.17\\
 & 15 &  5.25 & 319.70  & 481 & 2.46 & 1.63 & 3.34 & 39.50 & 53.85 & 28.17\\
 & 24 & 5.09 &  398.44  & 481 & 1.97 & 1.63 & 3.34 & 43.01 & 53.85 & 28.17\\
\hline 
\end{tabular}
\\
*All sizes are in kB
\caption{Values of various parameters for different images with size 786.6 kB at $\alpha = 46\%$}
\end{table*}
\end{center}

\begin{figure*}[!ht]
\begin{center}
   
  \subfloat[$k$ vs $C_r$]{%
      \includegraphics[width=0.27\textwidth]{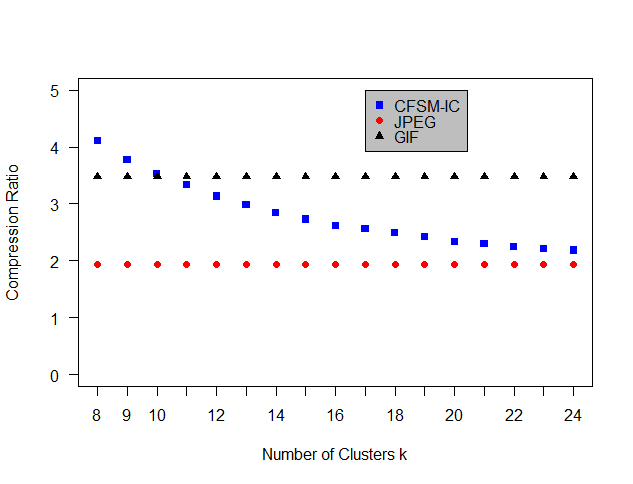}
    }
\hspace*{-0.4cm}
 \subfloat[$k$ vs PSNR]{%
      \includegraphics[width=0.27\textwidth]{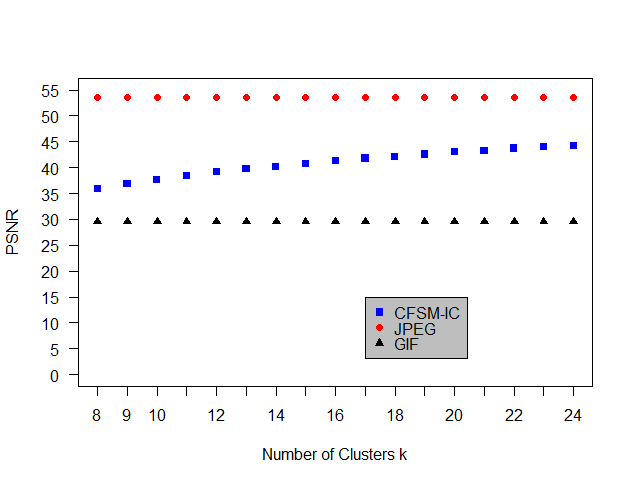}
    }
\hspace*{-0.4cm}
 \subfloat[$k$ vs Compression Time]{%
      \includegraphics[width=0.27\textwidth]{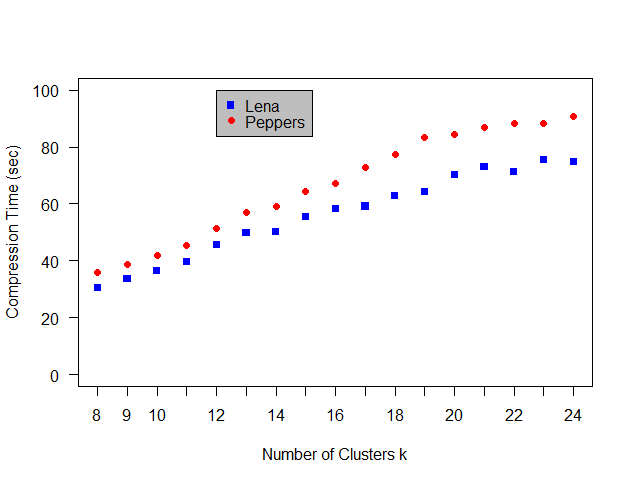}
    }
\hspace*{-0.4cm}
 \subfloat[ $\alpha$ vs $C_r$]{%
      \includegraphics[width=0.27\textwidth]{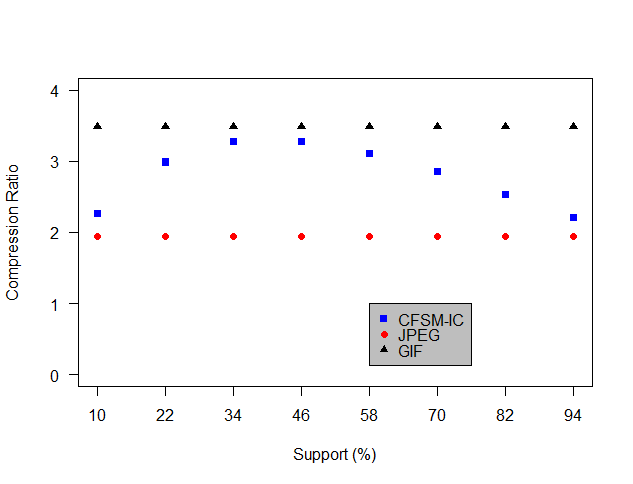}
    }
  
    \caption{Variation of $C_r$, PSNR, $k$ and $\alpha$ for Lena}
\end{center}
\end{figure*}

Image quality is the characteristic of an image that measures the degradation in image. The image quality is substantiated by using the metrics like PSNR(Peak Signal to Noise Ratio) and SSIM(Structural SIMilarity index).  PSNR is defined using the Mean Square Error (MSE) which measures the average of errors. MSE and PSNR are defined as follows.
\begin{center}
$MSE = \frac{1}{M N}\sum_{x=0}^{M-1}\sum_{y=0}^{N-1} [J(x, y) - J^{\hat{'}}(x, y)]^2$
\newline 

$PSNR = 20 \log_{10}\frac{MAX_J}{MSE}$
\end{center}

Here \textit{M, N} are the dimensions of the image indicating the number of rows and columns. \textit{J(x, y)} and $J^{'}(x, y)$ indicate the pixel value of the original image and the decompressed image at location\textit{(x, y)} respectively. $MAX_J$ denotes the maximum possible pixel value of the image \textit{J}. PSNR will be high if the average of errors MSE is low. SSIM measures the structural similarity between two images and is considered as a better parameter than PSNR, as PSNR doesn't account for visual perception.

\begin{figure*}[!ht]

    \subfloat[$\alpha$ vs Compression Time for Lena and Peppers  \label{hi}]{%
      \includegraphics[width=0.28\textwidth]{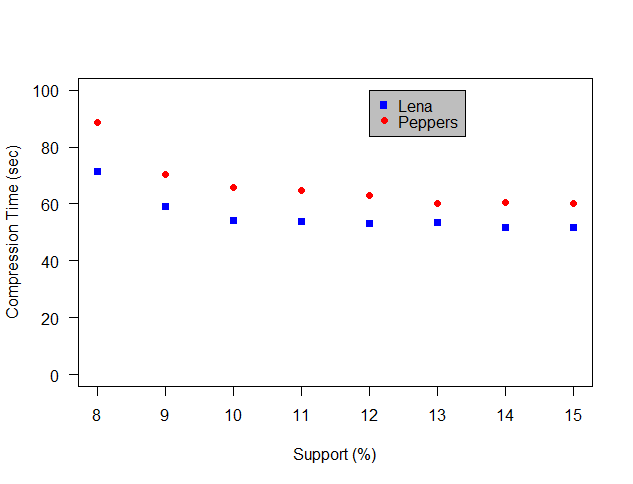}
    }
  \subfloat[Input Lena \label{hi}]{%
      \includegraphics[width=0.17\textwidth]{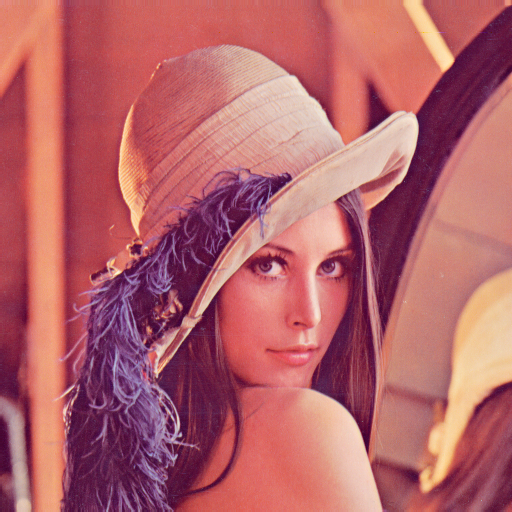}
    }
 \subfloat[$k=8$ \label{hello}]{%
      \includegraphics[width=0.17\textwidth]{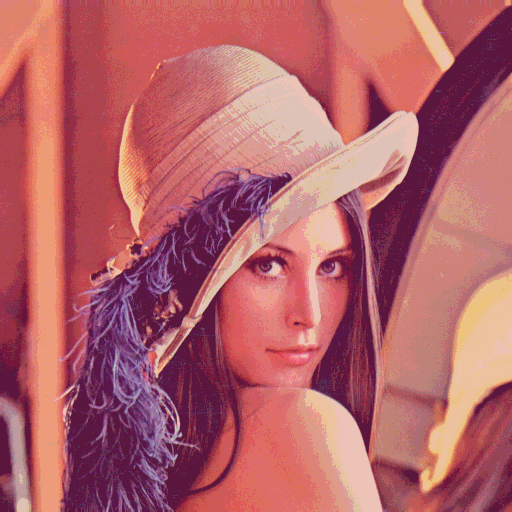}
    }
\hspace*{-0.2cm}
 \subfloat[$k=15$\label{hey}]{%
      \includegraphics[width=0.17\textwidth]{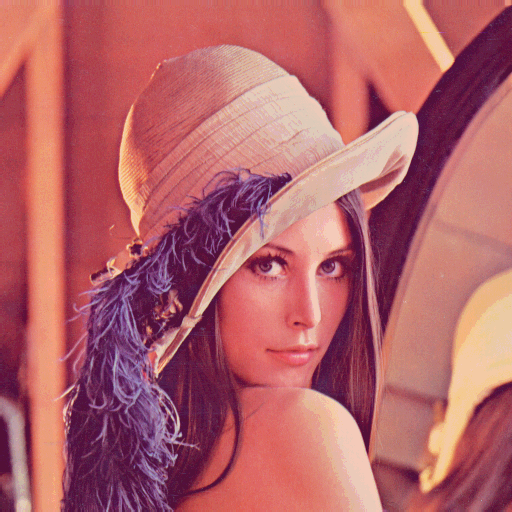}
    }
    \subfloat[$k=24$ \label{helloo}]{%
      \includegraphics[width=0.17\textwidth]{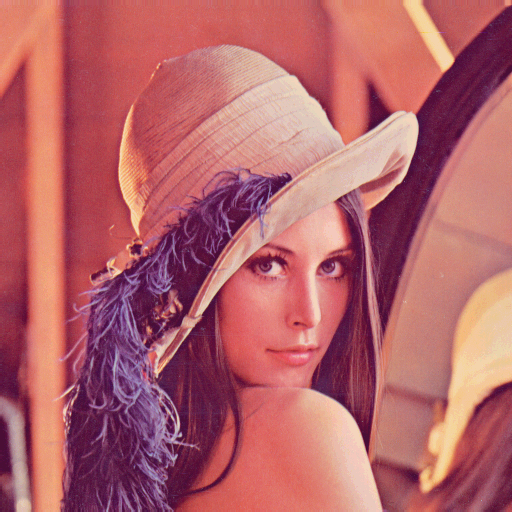}

    }
    \caption{(a) $\alpha$ vs Compression Time for Lena and Peppers (b) Lena Input Image (c), (d) and (e) are decompressed images at $k=8$, $k=15$ and $k=24$ respectively at $\alpha= 46\%$.}
    \label{fig:dummy}

  \end{figure*}

 The compressed size of the image $S_c$, denotes the overall sum of the Cluster Identifier table size, Code table size and Binary Encoded image size. Figure 2(a) shows the variation of the Compression ratio($C_r$) with the number of clusters \textit{k} for Lena image. It can be observed that as \textit{k} increases $C_r$ decreases. This is because an increase in the number of clusters decreases the number of lengthier closed frequent sequences thereby increasing the size of compressed image. For Lena image, the value of $C_r$ lies in [4.09, 2.18] for $k \in$ [8, 24]. For number of clusters $k \leq 10$, better compression ratio is achieved. A similar trend is observed for other standard images as well. On an average, there is a compression efficiency of 20 to 25\% in the selected range of $k$ for Lena and Peppers.

Image quality is affected solely by Clustering, as the original pixels are being replaced by their cluster identifiers which accounts for the loss in information. Figure 2(b) shows the effect of $k$ on the PSNR for Lena image. It can be observed that as $k$ increases, the PSNR increases and hence an image with significant quality is produced. This is due to grouping of pixels into more closely related clusters when $k$ is high, leading to a high PSNR. This leads to more number of cluster identifiers to represent the pixels which is not the case with low $k$ values. Also when $k \leq 10$, the compression is high compared to other values of $k$. The PSNR values of JPEG and GIF formats of Lena image are 53.6 at 404 $kB$ and 29.55 at 229 $kB$. The PSNR achieved through our algorithm varies from 36 to 44 for $k \in [8, 24]$. Our algorithm achieves around 45\% better quality in comparison to GIF for $k \geq 18$. For Peppers image, 40\% improvement in image quality is observed for $k\geq 18$. Similar trend can be seen in other images as well. The SSIM values of Lena are in between 0.94 and 0.99 whereas the JPEG format has an SSIM of 0.99 and GIF format has 0.78. The SSIM values of the images observed in our algorithm are close to JPEG and are better than GIF.

Increase in $k$ also results in increase in the compression time of Lena and Peppers which can be observed from Figure 2(c). It is because an increase in the number of clusters results in more searches to find the cluster identifier in the cluster identifier table and code table for the corresponding bit pattern. Figure 2(d) shows the effect of minimum support $\alpha$ on Compression Ratio $C_r$. Figure 3(b) refers to the original input image for Lena and figures 3(c)-(e) refer to the decompressed images at $\alpha=46\%$ and $k=8, 15, 24$ respectively for Lena. As $\alpha$ increases, the number of lengthier frequent sequences decreases which increases the size of the code tables. A high $C_r$ is achieved for $\alpha$ in the range $[40\%, 60\%]$. From 3(a), it can be depicted that $\alpha$ also has a significant effect on time. For lesser values of $\alpha$, more time is required to mine all the closed frequent sequences and thus an increase in compression time. For Lena image, at $k$ = 21 and $\alpha$ = 46, 21.60\% improvement in compression was observed in comparison to JPEG. For Peppers image, for the same values of $k$ and $\alpha$, 21.73\% improvement in compression is observed in comparison to JPEG. When $k \in [16, 22]$ and $\alpha \in [40,50]$, the quality of the decompressed image is observed to be better and other datasets also extends this trend.

\section{Conclusion}

We presented a novel, efficient and simple sequence mining based lossy image compression algorithm for images. We have proved using the experimental results that our method is good in mining the optimal pixel sequences and it still outperforms its competitors in compression ratio with an acceptable visual quality. Even then, our compression algorithm is considered as a prototype that needs further improvement. We intend to equip our proposed algorithm with better clustering technique to improve the quality of images using the properties of pixel neighbourhood. We would like to consider mining sub matrices instead of subsequences that would provide a significant improvement in the compression ratio.  Addition to that, we would aim to investigate with extensive simulations on large image size with more standard datasets, including the time to mine the sequence, which seems to be very promising. 

\bibliographystyle{IEEEtranS}
\bibliography{final_references}

\end{document}